\documentclass[journal=jacsat, manuscript=article,
amsmath,
amssymb,
]{achemso}
\usepackage[version=3]{mhchem} 
\usepackage{graphicx}
\usepackage{dcolumn}
\usepackage{bm}%
\usepackage{gensymb}
\usepackage{textgreek}
\usepackage{xcolor}
\usepackage{hyperref}
\usepackage{color,soul}
\usepackage{amsmath}
\usepackage{amssymb}
\usepackage[normalem]{ulem}

\author{Susumu Yada}
\affiliation[KTH Mech]{FLOW Centre, Dept. of Engineering Mechanics, Royal Institute of Technology (KTH), 100 44 Stockholm, Sweden}
\author{Ugis Lacis}
\affiliation[KTH Mech] {FLOW Centre, Dept. of Engineering Mechanics, Royal Institute of Technology (KTH), 100 44 Stockholm, Sweden}
\author{Wouter van der Wijngaart}
\affiliation[KTH MST]{Division of Micro and Nanosystems, Royal Institute of Technology (KTH), 100 44 Stockholm, Sweden}
\author{Fredrik Lundell}
\affiliation[KTH Mech]{FLOW Centre, Dept. of Engineering Mechanics, Royal Institute of Technology (KTH), 100 44 Stockholm, Sweden}
\author{Gustav Amberg}
\affiliation[KTH Mech]{FLOW Centre, Dept. of Engineering Mechanics, Royal Institute of Technology (KTH), 100 44 Stockholm, Sweden}
\alsoaffiliation[Sodertorn]{S\"odert\"orn University, 141 89 Stockholm, Sweden}
\author{Shervin Bagheri}
\affiliation[KTH Mech]{FLOW Centre, Dept. of Engineering Mechanics, Royal Institute of Technology (KTH), 100 44 Stockholm, Sweden}%
\email{shervin@mech.kth.se}

\title[An \textsf{achemso} demo]
  {Droplet impact on asymmetric hydrophobic microstructures}

\abbreviations{IR,NMR,UV}

\begin{document}

\begin{tocentry}
    \begin{center}
    \includegraphics[width=5cm]{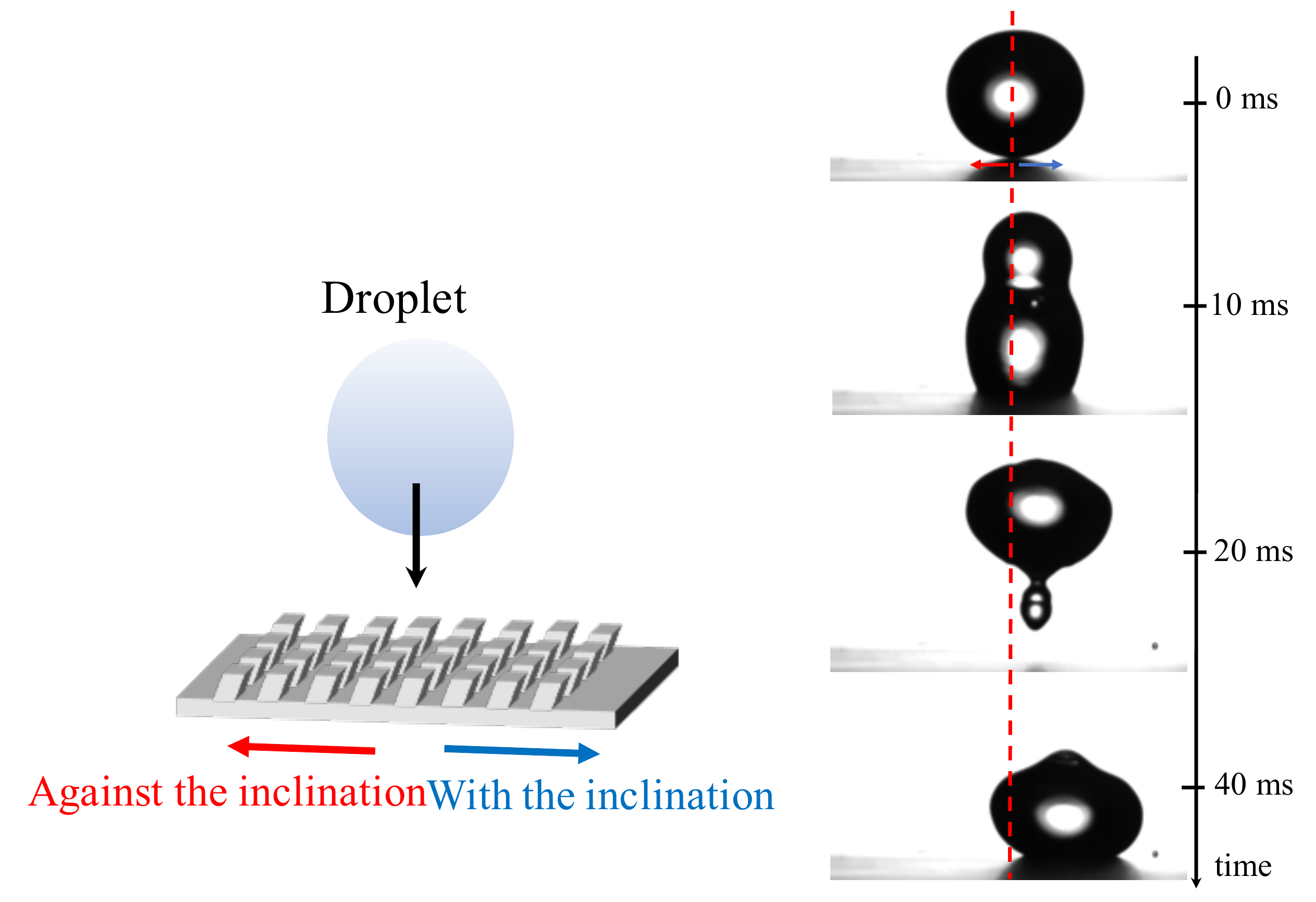}
    \end{center}
\end{tocentry}

\begin{abstract}
 Textured hydrophobic surfaces that repel liquid droplets unidirectionally are found in nature such as butterfly wings and ryegrass leaves and are also essential in technological processes such as self-cleaning and anti-icing.
However, droplet impact on such surfaces is not fully understood.
Here, we study, using a high-speed camera, droplet impact on surfaces with inclined micropillars.
We observed directional rebound at high impact speeds on surfaces with dense arrays of pillars.
We attribute this asymmetry to the difference in wetting behavior of the structure sidewalls, causing slower retraction of the contact line in the direction against the inclination compared to with the inclination. 
The experimental observations are complemented with numerical simulations to elucidate the detailed movement of the drops over the pillars.
These insights improve our understanding of droplet impact on hydrophobic microstructures and may be a useful for designing structured surfaces for controlling droplet mobility. 
\end{abstract}

\section{Introduction}
Droplet deposition and impact are important in applications such as spray coating and cooling \cite{Dykhuizen1994, Josserand2016}, pesticide deposition \cite{Bergeron2000, Liu2017NRM}, and inkjet printing \cite{AttingerJHT2000, Minemawari2011}.
In particular, droplet impact involves complex fluid motions including splashing \cite{MUNDO1995151, XuPRL2005, josserand_lemoyne_troeger_zaleski_2005, DriscollPRL2011, RibouxPRL2014}, the formation of a thin gas layer between the droplet and the surface \cite{MandrePRL2009,BouwhuisPRL2012, RoelandSM2014, VisserSM2015}, and droplet rebound on superhydrophobic surfaces \cite{BartoloEPL_2006, Reyssat_2006, TsaiLangmuir2009, Gauthier2015,  PATIL_ETFS2016195}.
Theoretical \cite{RoismanPRSLA2002, Attane2007PoF, WildemanJFM2016, EggersZaleskiPoF2010}, numerical \cite{EggersZaleskiPoF2010, SchrollPRL2010, WildemanJFM2016, WANGJNFM2017}, and experimental investigations \cite{clanet_2004, KANNAN2008694,TsaiLangmuir2009, LaanPRA2014, LeeLangmuir2016, LIN201886, LeeJFM2016, Yada_langmuir_2021} of droplet impact have also highlighted the fingering of spreading front and the scaling laws for maximum deformation \cite{Yarin2006, Josserand2016, Cheng_Gordillo_ARFM_2022}.

Despite a long-time understanding of droplet rebound from textured hydrophobic surfaces after impact, only recently we understand the coupling between the droplet dynamics and surface microstructure.
These surfaces trap air underneath droplets (i.e. the Cassie wetting state), rendering the surface significantly more hydrophobic. 
The robustness of the air cavity and the resulting droplet behavior has been studied in terms of impalement pressure \cite{Reyssat_2006, BartoloEPL_2006, QuereARMR2008, PATIL_ETFS2016195}. When the fluid pressure exceeds a critical pressure on a structured surface, the Cassie-Wenzel wetting transition occurs and droplets cease to rebound.

Asymmetric hydrophobic microstructures are often exploited by natural species, such as butterfly wings \cite{Li_Langmuir2018} and ryegrass leaves \cite{GuoSM2012} where they assist liquid roll-off. 
Surfaces with asymmetric ratchets and spikes allow directing a droplet in a desired direction and such anisotropic surfaces are useful particularly in self-cleaning, water harvesting \cite{YuHuang_Advmat_2016}, and cell-directing \cite{GuoSM2012, Malvadkar2010,Liu_Angewandte_2016}. 
Here, hydrophobic surface properties are advantageous to increase the mobility of a droplet.
On such hydrophobic surfaces, upon impact, droplets bounce off towards the direction in which the surface structures are oriented \cite{Lee_ACSn2018, Li_Langmuir2018, Li_RESA_2017, Lu_Langmuir_2020, PLi_ACSAMI_2021}. 
However, the detailed mechanisms of bouncing 
are not fully understood. In particular, the influence of the surface geometry, i.e., the pitch and height of structural features, and the impact velocity remain to be fully elucidated.

Here, we study droplet impact on asymmetric microstructures experimentally and numerically.
We observe the distinct influence of surface geometry and impact velocity on impact behaviour. 
Moreover, we measure the trajectories of bouncing droplets and investigate the conditions for directional rebound. 
We observe and discuss differences in receding speeds of the contact line in the direction with the inclination and against the inclination.

\section{Materials and Methods}
\subsection{Experimental set up}
\begin{figure}
    \centering
    \includegraphics[width=0.9\textwidth]{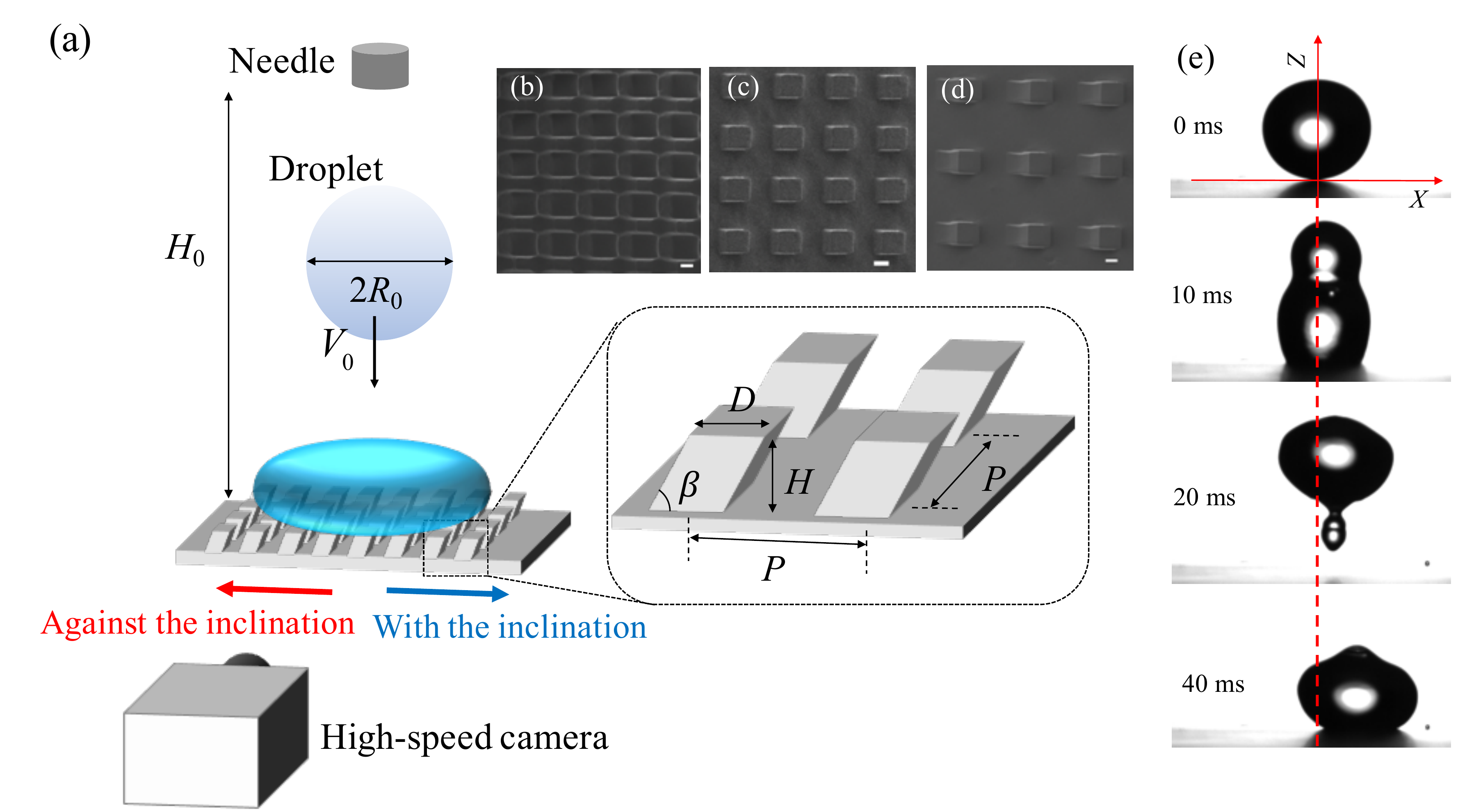}
    \includegraphics[width=0.9\textwidth]{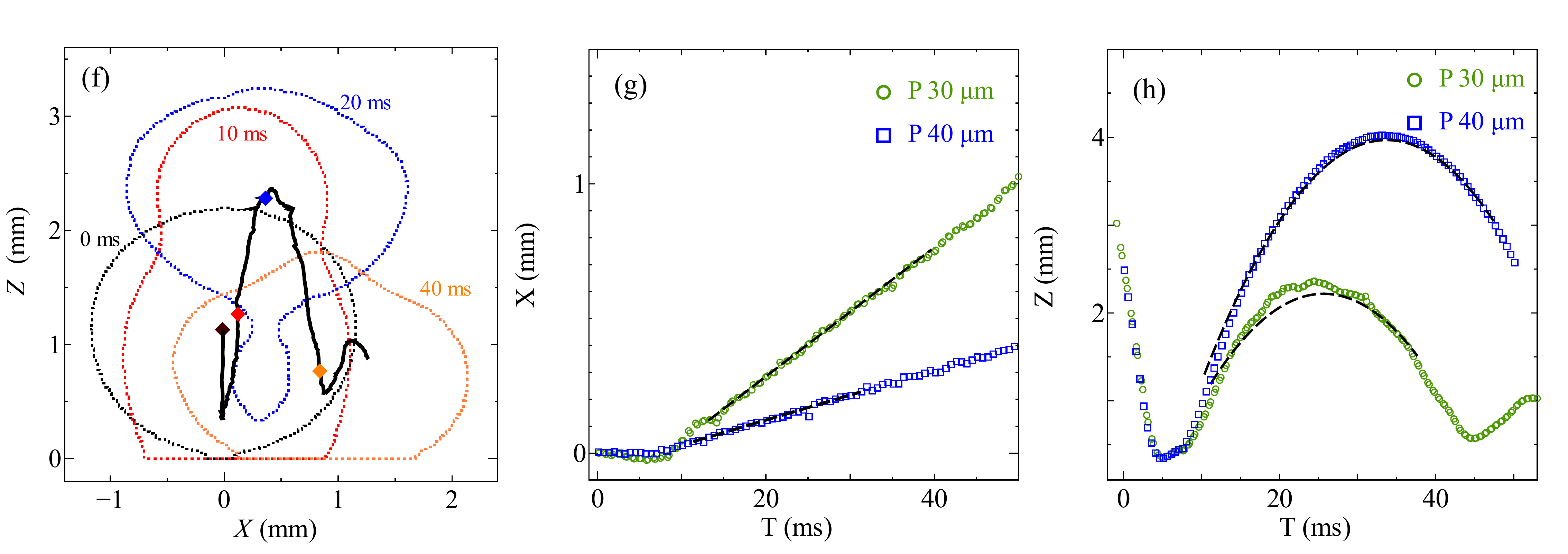}
    \caption{
    (a) Schematic description of the droplet impact experiment.
    (b-d) Scanning electron microscopy image of the inclined microstructure with (b) $P= 30~\mathrm{\mu m}$, (c) $P= 40~\mathrm{\mu m}$, (d) $P= 60~\mathrm{\mu m}$. The scale bar indicates 10~$\mathrm{\mu}$m.
    (e) Selected snapshots from experiments ($P = 30~\mathrm{\mu m}$ and $V_0 = 0.56$~m/s). The surface structures are inclined to the right.  
    (f-h) Procedure to estimate rebound velocity.
    (f) Captured droplet shape (dotted lines) and the trajectory of center of mass (black solid line) from (e). The positive $X$ indicates the horizontal direction with the inclination of the pillars and $Z$ is the vertical displacement from the substrate.
   (g, h) The horizontal and vertical position of the center of mass as a function of time.
    Dash black lines in (g, h) are ballistic trajectories (Eqs.~\ref{eq:X}, \ref{eq:Z}) with fitted $V_x$ and $V_z$. A typical trajectory for $P= 40~\mathrm{\mu m}$ and $V_0 = 0.56$~m/s is also shown in (g, h).
    }
    \label{fig:First}
\end{figure}

The impact of liquid droplets is observed with a high-speed camera (speedsense, Dantec dynamics) at a frame rate of 6000-8000~$\mathrm{s^{-1}}$ with a spatial resolution of 15~$\mathrm{\mu m}$. The schematics of the experimental setup are shown in Fig.~\ref{fig:First}(a).
A liquid droplet is formed on the tip of a needle with an outer diameter of 0.31 mm (Hamilton, Gauge 30, point style 3) at a height $H_0$ from the surfaces. The liquid is pumped by a syringe pump (Cetoni, neMESYS 1000N) at a small flow rate (0.10~$\mathrm{\mu L/s}$) and the droplet pinches off from the needle with the constant initial radius $R_0$ = 1.14 $\pm$ 0.02~mm. The droplet is accelerated by gravity and hits the substrate with an impact velocity $V_0$. 
The impact velocities are varied by changing the distance from the substrate to the needle $H_0$. The impact velocity is estimated from the images just before the droplet hits the substrate. The captured images are shown in Fig.~\ref{fig:First}(e).
The height $H_0$ is varied from 5~mm to 85~mm, which leads to impact velocities $V_0$ from 0.25~m/s to 1.3~m/s (Table \ref{tab:Height}). 

 The liquid employed in this study is deionized water. The surface tension of water $\sigma$ is measured to be 0.072~mN/m with TD 2 tensiometer (LAUDA).
 In this study, we focus on the droplet motion in the direction of the inclination of the pillars.

\begin{table}
    \centering
    \begin{tabular}{lcccccccccc}
    \hline
         $H_0$(mm)  & 5 & 10 & 15 & 20  & 25 & 40 & 60 & 85 \\
         \hline
        $V_{0}$(m/s) & 0.25 & 0.38 & 0.50 & 0.56 & 0.64 & 0.84 & 1.1 & 1.3 \\
        $We$ & 0.9 & 2.3 & 3.8 & 5.1 & 7.3 & 10.8 & 16.8 & 25.8 \\
    \hline
    \end{tabular}
    \caption{List of the heights $H_{0}$, the impact velocities $V_0$, and Weber number $We =\rho R_0 V_0^2 /\sigma$.}
    \label{tab:Height}
\end{table}
\subsection{Surface preparation}
The substrates studied are made from Ostemer 220 (Mercene Labs, Sweden), Off-Stoichiometry-Thiol-Ene (OSTE) resin \cite{Carlborg2011, Yada_SM}. The resin is suitable for fabricating inclined micro patterns by exposing slanted collimated ultraviolet light. 
The surfaces are prepared in three steps. 
First, a base OSTE layer is prepared on a smooth plastic film.
Secondly, inclined micropillars are developed on the base layer by exposing slanted ultraviolet light through a patterned photomask. After cleaning uncured OSTE in an acetone bath, hydrophobic surface modification using 1\% w/w fluorinated methacrylate (3,3,4,4,5,5,6,6,7,7,8,8,9,9,10,10,10-Heptadecafluorodecyl methacrylate, Sigma Aldrich) solution in 2-Propanol with 0.05\% benzophenone (Sigma-Aldrich) initiator is applied. 
Surface structures are characterized with scanning electron microscopy (see Fig.~\ref{fig:First}b-d) and the inclination of the pillars $\beta$ is 60 degrees. 

The equilibrium contact angles of deionized water are reported in Table.~\ref{tab:surface}. The advancing and receding contact angles are measured using the sessile drop method \cite{EralCPS2013, Korhonen_Langmuir_2013}.
A droplet with the initial volume of $5~\mathrm{\mu L}$ is deposited on the surface and it is pumped through the needle at a flow rate of 0.1~$\mathrm{\mu}$L/s to measure advancing contact angle. For the receding angle measurements, the initial volume is set to $30~\mathrm{\mu L}$ to perform reliable measurements \cite{Korhonen_Langmuir_2013} and the droplet is drained at a flow rate of 0.1~$\mathrm{\mu}$L/s.
The average contact angle for 5 seconds after the contact line starts to move is defined as the advancing (receding) contact angle.

\subsection{Rebound velocity estimation}
To investigate the influence of the surface structure and the impact velocity on rebound behaviors, the trajectory of the droplet is calculated.
The trajectory of the center of mass is obtained by extracting the surface contour from the images (Fig.~\ref{fig:First}e, f).
Assuming ballistic trajectory after the impact, the horizontal and vertical positions $X, Z$ are described as a function of time $T$,
\begin{equation}
X(T)=X_0+V_x (T-T_0),      
\label{eq:X}
\end{equation}
\begin{equation}
Z(T)=Z_0+V_z (T-T_0) - g(T-T_0)^2 /2,  
\label{eq:Z}
\end{equation}
where The positive $X$ indicates the horizontal direction with the inclination of the pillars and $Z$ is the vertical displacement from the substrate.
Here, $V_x, V_z$ are horizontal and vertical velocities, $g$ is the gravitational acceleration, and $X_0$ and $Z_0$ are the horizontal and vertical positions at $T=T_0$. Equations \ref{eq:X},\ref{eq:Z} describe the trajectory well when $V_x, V_z $ are fitted (see the dash lines in Fig.~\ref{fig:First}g, h). 
By performing the fitting procedure, $V_x$ and $V_z$ are estimated as a function of the impact velocity. Here, $X_0, T_0$ are set so that $Z_0$ = 1.1 $R_0$ for all configurations.
\begin{table}
    \centering
    \scalebox{0.9}{
    \begin{tabular}{lcccccccccc}
    \hline
          & $D~(\mathrm{\mathrm{\mu m}})$ & $P~(\mathrm{\mu m})$ & $H~(\mathrm{\mu m})$ & $\theta_e$ (deg) & $\theta_{a-A}$ (deg) & $\theta_{a-W}$ (deg) & $\theta_{r-A}$ (deg) & $\theta_{r-W}$ (deg)  \\
    \hline
     Flat & - & - & - & 112 $\pm$ 2 & 121 $\pm$ 3 & - &73 $\pm$ 3 & -\\
      P30 & 20 & 30 & 20 & 133 $\pm$ 3 & 150 $\pm$ 2 & 146 $\pm$ 4 & 84 $\pm$ 8 & 100 $\pm$ 4  \\
    P40 & 20 & 40 & 20 & 146 $\pm$ 1 & 147 $\pm$ 6 & 147$\pm$ 5 & 115 $\pm$ 5 & 117$\pm$ 5 \\
     P60 & 20 & 60 & 20 & 109 $\pm$ 3 & 112 $\pm$ 5& 117 $\pm$ 5 & 62 $\pm$ 5 & 58 $\pm$ 5\\
     \hline
    \end{tabular}
    }
    \caption{List of the surfaces. $D, P, H$ are the width, pitch, and height of the pillars. $\theta_e$ is the equilibrium contact angle. $\theta_{a-A}$, $\theta_{a-W}$ , $\theta_{r-A}$ , $\theta_{r-W}$ are advancing/receding contact angle in the direction against the inclination and with the inclination, respectively. The advancing/receding contact angles in the direction against the inclination and with the inclination on the flat surface are identical. 
    }
    \label{tab:surface}
\end{table}


\section{Results}

\subsection{Bouncing regimes}

\begin{figure}
    \centering
    \includegraphics[width=0.9\textwidth]{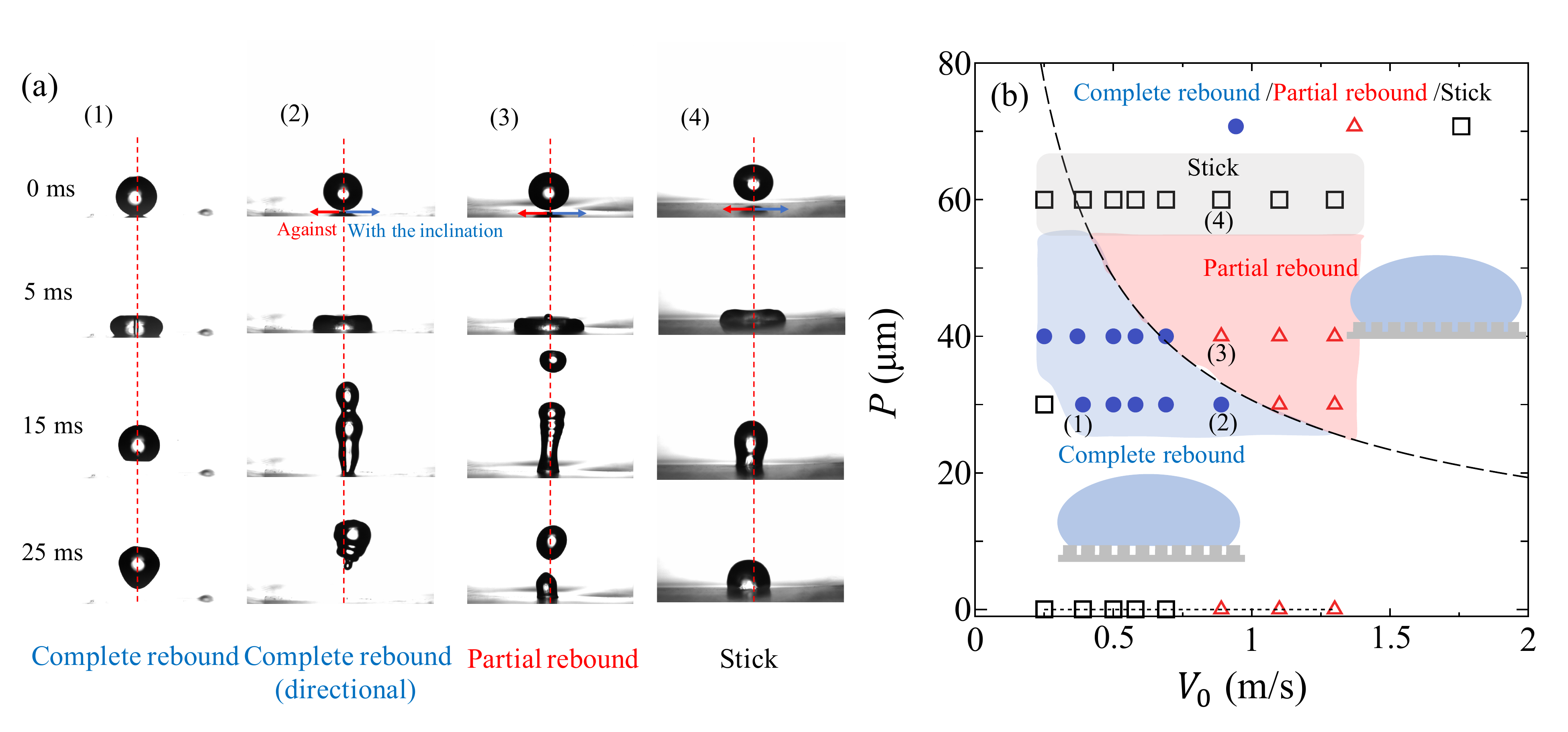}
    \caption{
    (a) Selected snapshots from the experiments. (1) $V_0$ = 0.38~m/s on $P=30~\mathrm{\mu m}$  (2) $V_0$ = 0.84~m/s on $P=30~\mathrm{\mu m}$ (3) $V_0$ = 0.84~m/s on $P=40~\mathrm{\mu m}$ (4) $V_0$ = 0.84~m/s on $P=60~\mathrm{\mu m}$. The surfaces structures are inclined to the right.
  (b) Impact velocity-pitch map for different behavior after droplet impact. Blue, red, black marks indicate complete rebound, partial rebound, and stick behavior, respectively. $P$ = 0 indicates the flat surfaces. The dashed curve describes the semi-quantitative model for an array of straight pillars by Bartolo et al. \cite{BartoloEPL_2006} $V_{c} = \sqrt{\sigma H D/\rho P^3}$. The inset shows the schematic for the wetting transition. The numbers (1-4) correspond to the snapshots in (a).
    }
    \label{fig:Merged}
\end{figure}

Figure~\ref{fig:Merged} shows series of images of a water droplet spreading after impact. We observe that the pitch between the pillars $P$ and the impact velocity $V_0$ determine the droplet behavior. 

Three distinctive behaviors are observed. First, the droplet completely rebounds from the surface (1,2 in Fig.~\ref{fig:Merged}a). Moreover, the droplet rebounds to the direction with the inclination on $P = 30~\mathrm{\mu m}$ and at high $V_0$ (case (2) in Fig.~\ref{fig:Merged}a).
Secondly, the droplet breaks up and part of the droplet remains deposited on the surface while the other part bounces up (3 in Fig.~\ref{fig:Merged}a).
Finally,  the droplet does not bounce and sticks to the surface (4 in Fig.~\ref{fig:Merged}a).
We refer to the three configurations as "Complete rebound", "Partial rebound", and "Stick". 
Figure~\ref{fig:Merged}(b) shows the pitch-impact velocity parameter map with the "Complete rebound", "Partial rebound", and "Stick" regions are indicated.

The Cassie-Wenzel transition is responsible for the different behaviors. For "Complete rebound" situations, the grooves between the posts are not wetted and air is trapped underneath the droplet (Cassie state). On the other hand, for "Partial rebound" and "Stick" cases, the grooves are partially or fully penetrated by the liquid (Wenzel state). 
A semi-quantitative model to account for the Cassie to Wenzel transition on an array of pillars was proposed by Bartolo \textit{et al.} \cite{BartoloEPL_2006} The model estimates the critical impalement pressure on a structured surface. When the hydrodynamic pressure over the surface exceeds the critical pressure, the liquid-air interface makes contact with the basal surface of the substrate and the liquid penetrates into the grooves. 
Above the critical pressure, the Cassie-Wenzel wetting transition occurs, which also corresponds to the transition from bouncing to non-bouncing.
The model estimates the critical pressure as $p_c \sim \sigma H D/2P^3$ for dense arrays of straight pillars \cite{BartoloEPL_2006}, where $H$ and $D$ are the height and width of the pillars, respectively. In the instant of droplet impact, the hydrodynamic pressure is $p_d \sim \rho V_0^2 /2$ where $\rho$ is the density of the liquid. The balance $p_c \sim p_d$ gives the critical impact velocity for the pitch $P$ as $V_c \sim \sqrt{\sigma H D/\rho P^3}$. The dashed curve in Fig.~\ref{fig:Merged}(b) depicts this critical value. 
We observe that the critical curve separates the complete rebound regime and partial rebound regime reasonably well also for inclined pillars. Beyond the critical impact velocity, "Stick" and "Partial rebound" are observed.

\subsection{Rebound velocity}

As seen in the previous sections, the rebound behavior in the horizontal direction depends on the surface structure and the impact velocity. 
This section investigates the directional behavior within the rebound regime.

\begin{figure}[t!]
    \centering
    \includegraphics[width=0.63\textwidth]{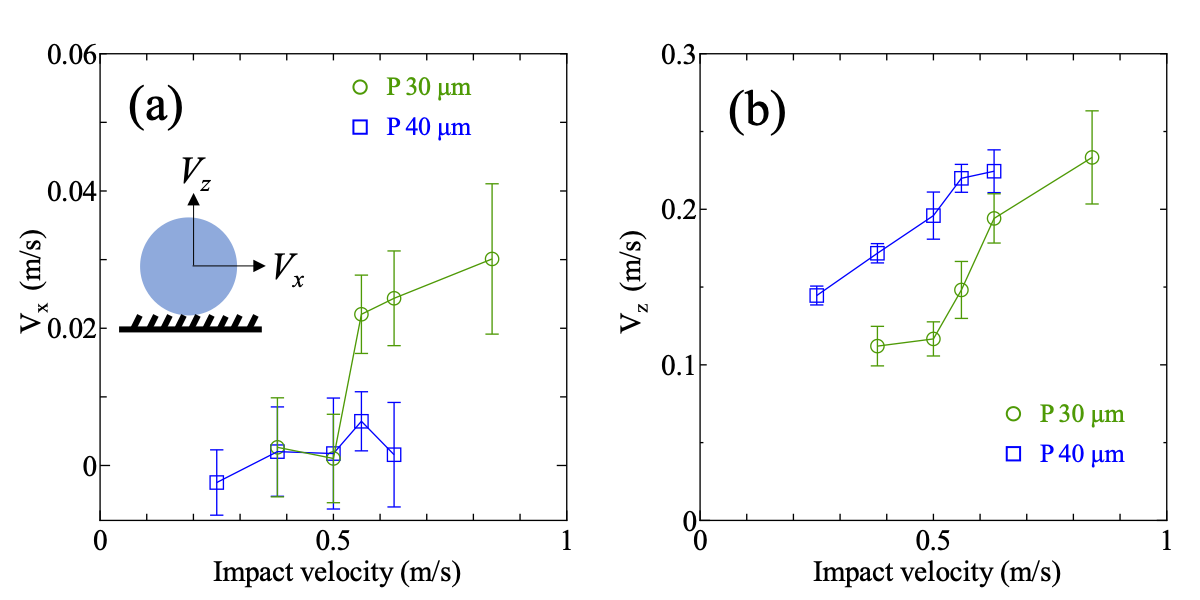}
     \includegraphics[width=0.315\textwidth]{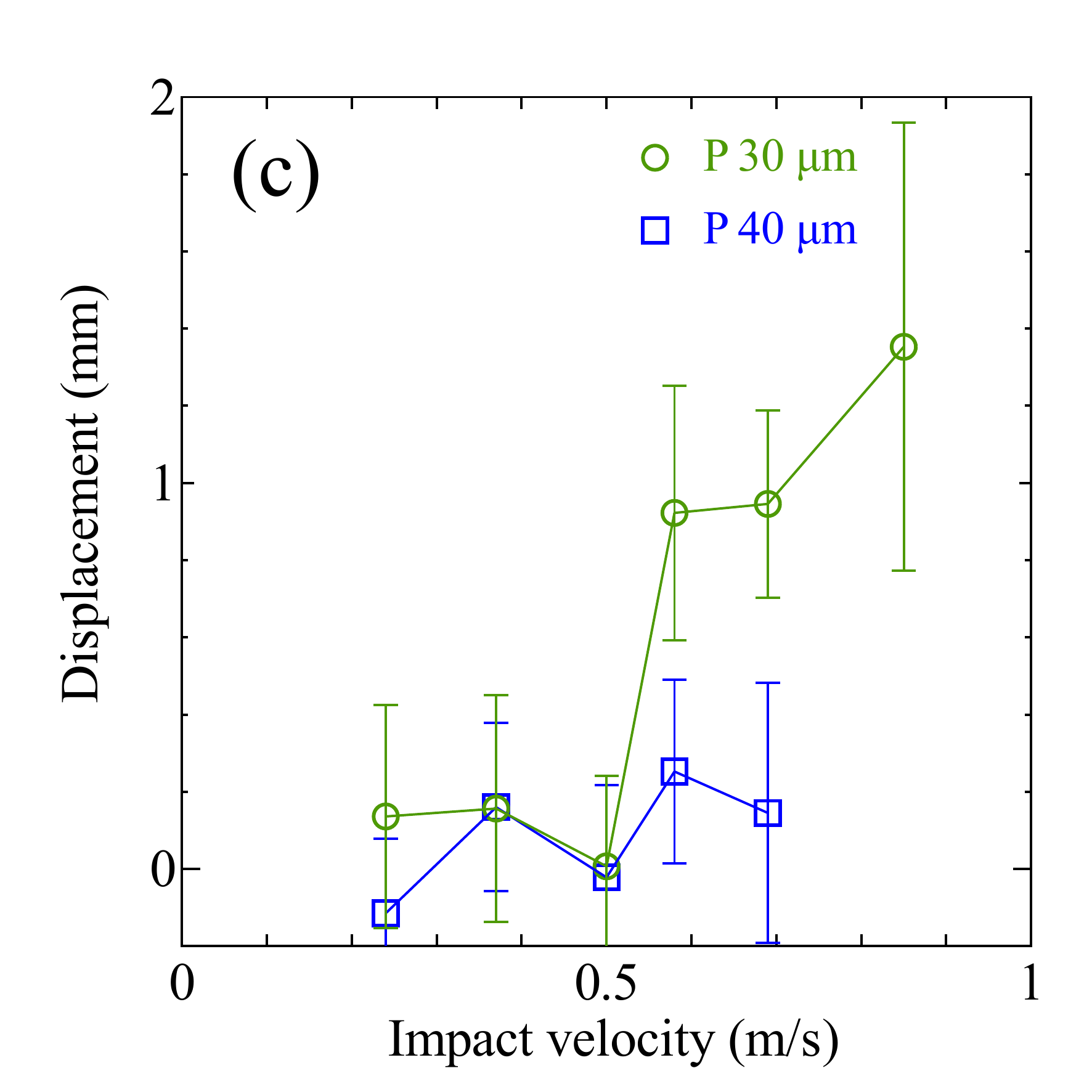}
    \caption{
The influence of impact velocity on droplet rebound velocity. 
    (a) Horizontal rebound velocity $V_x$. (b) Vertical rebound velocity $V_z$. (c) The terminal horizontal displacement for different impact velocity. 
    Error bars indicate standard deviations. The data are averages over more than 8 separate measurements. 
    }
    \label{fig:VxVz}
\end{figure}

 Figure~\ref{fig:VxVz}(a) shows $V_x$ at different impact velocities.
The horizontal velocity $V_x$ is negligibly small for low impact velocity ($<$ 0.5~m/s) and increases to $\sim$ 0.03~m/s with the impact velocity for $P= 30~\mathrm{\mu m}$. For $P= 40~\mathrm{\mu m}$, $V_x$ remains small even for the highest impact speed.

The vertical rebound velocity $V_z$ in Fig.~\ref{fig:VxVz}(b) increases with the impact velocity up to $\sim$ 0.25~m/s. Larger $V_z$ is observed for $P = 40~\mathrm{\mu m}$ compared to $P = 30~\mathrm{\mu m}$. This is likely because of the higher level of hydrophobicity, which is indicated by the larger equilibrium contact angle on $P = 40~\mathrm{\mu m}$ (see Table.~\ref{tab:surface}). As a result, the droplet moves in the direction with the inclination up to 1.3~mm (Fig.\ref{fig:VxVz}c). The directional displacement is observed only for $V_0 >$ 0.5~m/s and on $P = 30~\mathrm{\mu m}$.
This is similar to the observation made by Li \textit{et al.} \cite{PLi_ACSAMI_2021}. They also observed a larger horizontal displacement on arrays of inclined cones with a smaller spacing. 

\begin{figure}[t]
    \centering
   \includegraphics[width=0.5\textwidth]{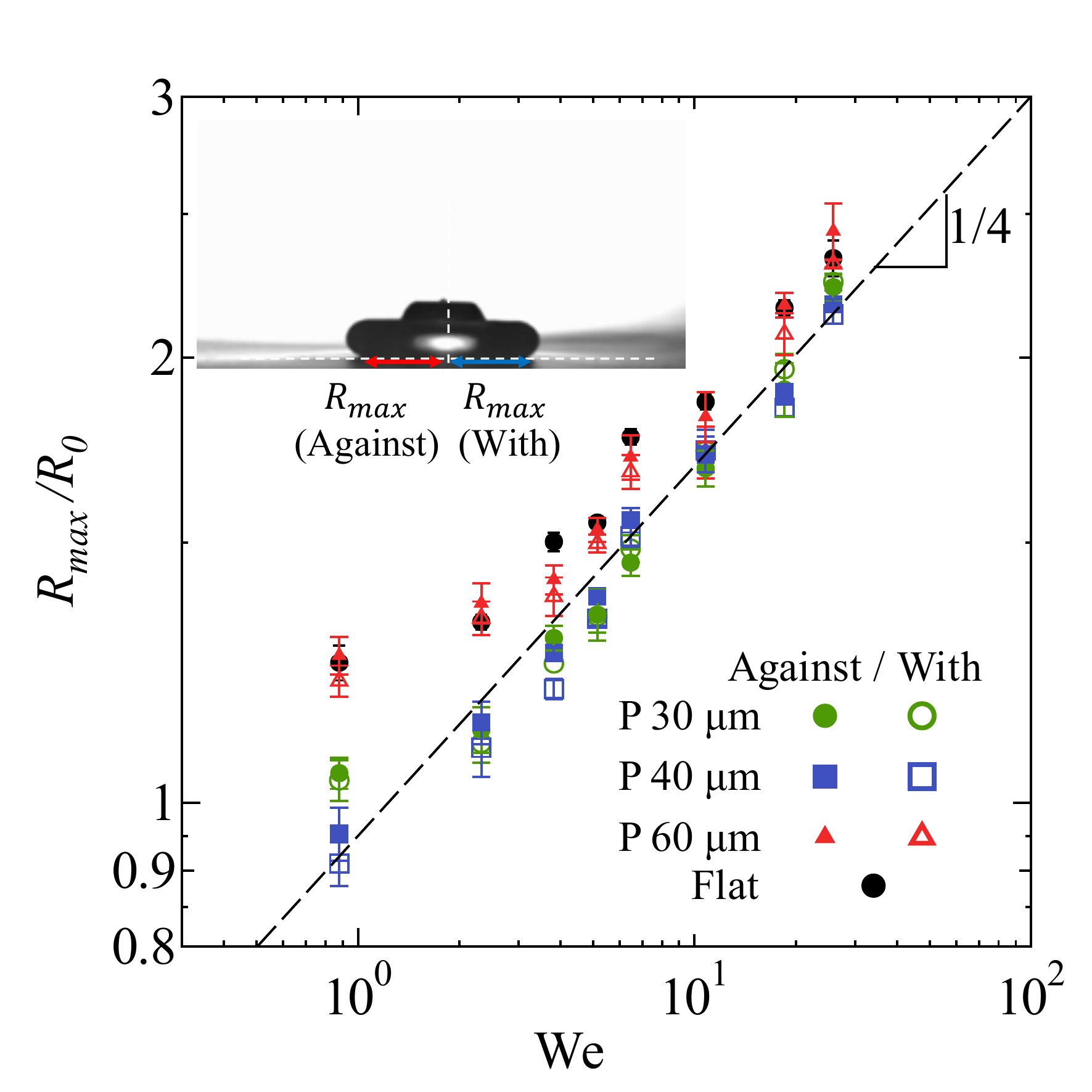}
    \caption{Normalized maximum contact radius as a function of Weber number $We= \rho R_0 V_0^2 /\sigma$. The dashed line indicates $R_{max} \propto We^{1/4}$. The inset describes the definition of $R_{max}$. The surface in the inset is oriented to the right. }
    \label{fig:Rmax}
\end{figure}

It is noticeable that the expansion phase until the droplet reaches the maximum deformation is symmetric on the inclined hydrophobic pillars (see the snapshots at 5~ms in Fig.~\ref{fig:Merged}).
This is further quantified in Fig.~\ref{fig:Rmax}, where the maximum contact radius of the droplet $R_{max}/R_0 $ is shown as a function of the Weber number $W e= \rho R_0 V_0^2 /\sigma$. The maximum contact radii in the direction with the inclination and against the inclination are similar for all surfaces.
The maximum contact radius for $P = 30~\mathrm{\mu m}$ and $P = 40~\mathrm{\mu m}$ are slightly smaller than for the flat surface, while it is nearly the same as the flat surface for $P = 60~\mathrm{\mu m}$. 
Furthermore, the maximum contact radius follows the well-known relation \cite{clanet_2004, LIN201886} $R_{max} \propto We^{1/4}$. 
This is consistent with previous studies with low Ohnesorge number $Oh = \mu/\sqrt{\rho R_0 \sigma}$ where $\mu$ is the liquid viscosity, while more viscous fluids exhibit a smaller exponent ($\sim$ 1/6) \cite{LIN201886, clanet_2004, Attane2007PoF}. The Ohnesorge number in this study is $3.5 \times 10^{-3}$, which is reasonably low. 

Contrary to the first expansion, the retraction immediately after the initial expansion can be asymmetric. The asymmetric retraction is responsible for the observed asymmetric bouncing. 
The retraction is governed by how the contact line detaches from the surface structures.

\begin{figure}[t]
    \centering
     \includegraphics[width=0.4\textwidth]{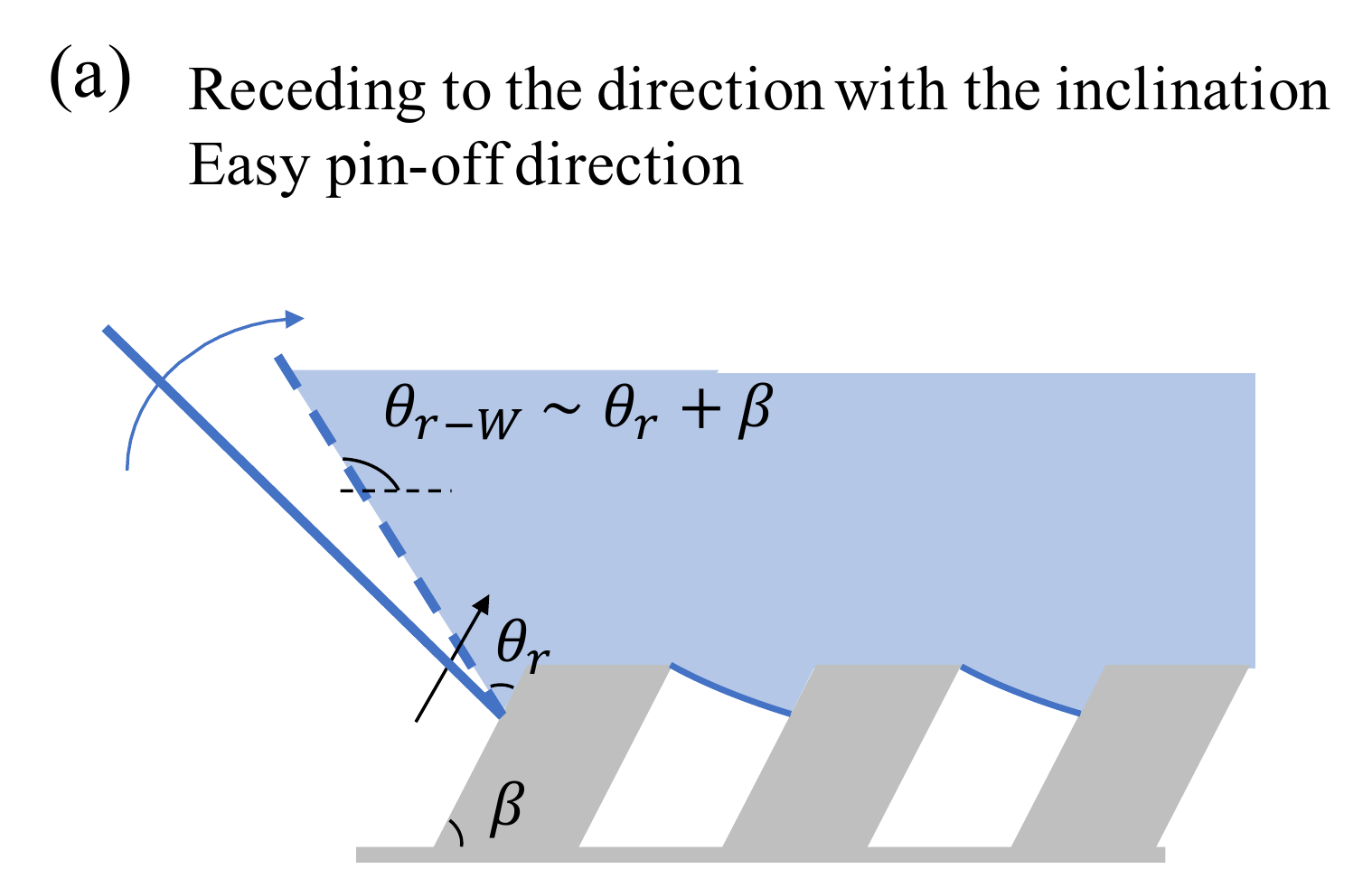}
      \includegraphics[width=0.4\textwidth]{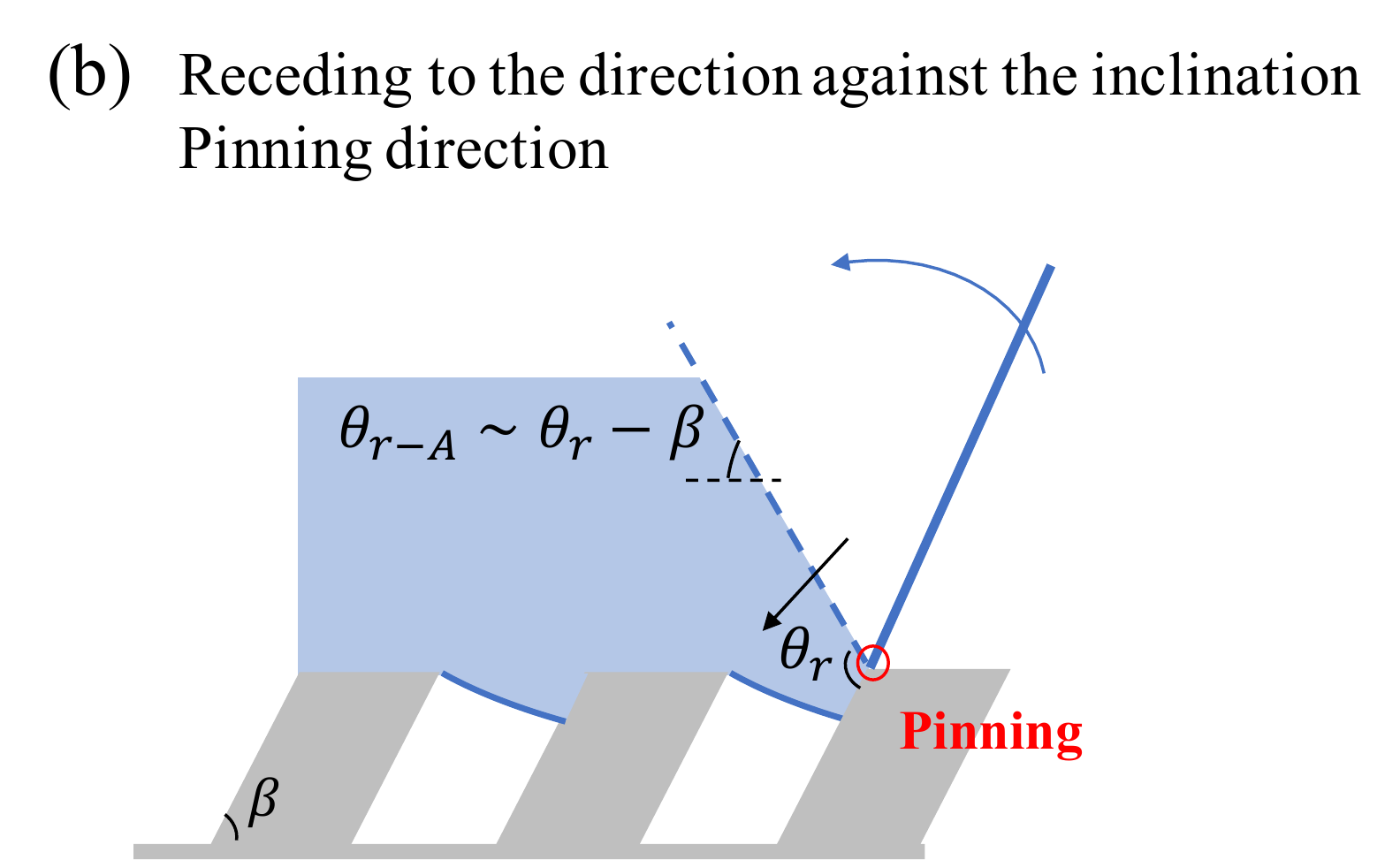}
    \caption{ Schematic models of the receding contact lines (a) in the direction with the inclination (b) in the direction against the inclination.
    }
    \label{fig:schematics}
\end{figure}

The underlying mechanism of the asymmetric receding speed is in the wetting of the asymmetric microstructure.
Figure~\ref{fig:schematics} shows the schematic model of the receding contact line on the asymmetric microstructure. 
The key factor is that a part of the inclined sidewall is wetted.
The wetting on asymmetric microstructured surfaces was also illustrated by Guo \textit{et al.} \cite{GuoSM2012} and Malvadkar \textit{et al.} \cite{Malvadkar2010} for rolling-off droplets.

The contact line recedes when the local contact angle decreases to the intrinsic receding contact angle, $\theta_r$= 73 degrees.
When the contact line recedes in the direction with the inclination on the inclined wall (from left to right in Fig.~\ref{fig:schematics}a), the apparent receding angle $\theta_{r-W}$ is $\theta_r + \beta \sim$ 133 degrees.
Therefore, the contact line smoothly recedes on the sidewall. 
On the other hand, the apparent receding angle in the direction against the inclination ($\theta_{r-W}$) -- when the contact line moves down on the sidewall -- should be $\theta_r - \beta$, i.e.~as small as 13 degrees (Fig.~\ref{fig:schematics}b). The contact line is then pinned at the obtuse corner until the liquid detaches from the sidewall. 
This pinning delays the receding in the direction against the inclination. 
Note that the liquid inertia helps the interface detach from the sidewall, so the apparent receding angle in the experiments is not as small as $\theta_r - \beta $. 

\subsection{Numerical simulations of droplet impact}

\begin{figure}
    \centering
    \includegraphics[width=0.9\textwidth]{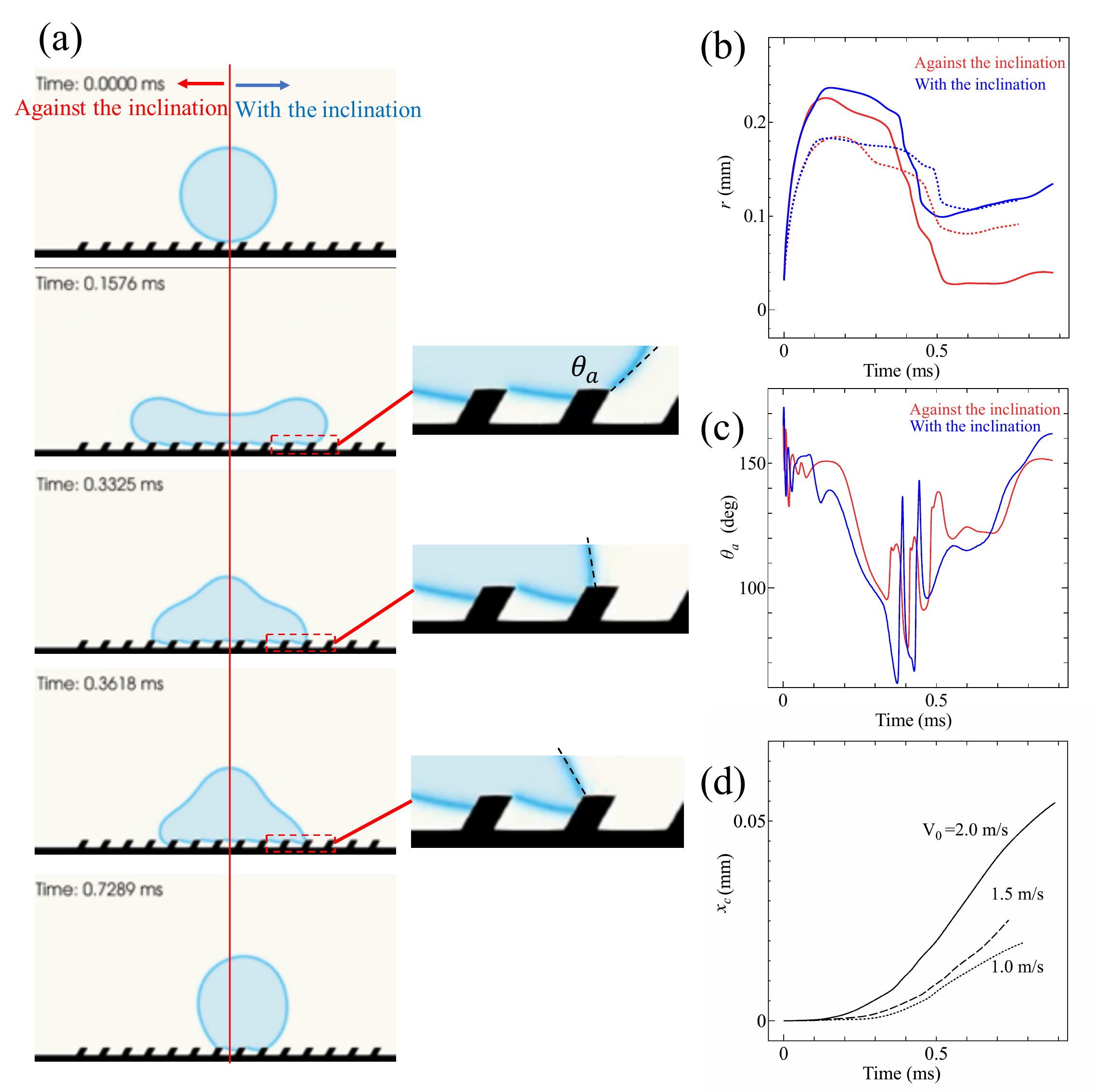}
    \caption{
    Simulations of a droplet impacting on an asymmetric hydrophobic microstructure (a) Selected pictures from the simulation. The inset provides a magnified picture near the contact line. The impact velocity in (a) is 2~m/s. (b) The contact radii from the initial center of the droplet. The solid and dot lines in (b) indicate $V_0$= 2~m/s and 1~m/s, respectively.
    (c) The apparent contact angle $\theta_a$. The impact velocity in (c) is 2~m/s. (d) The center of mass $x_c$. The positive $x_c$ indicates the horizontal direction with the inclination. The solid, dash, dot lines in (d) indicate $V_0$= 2~m/s, 1.5~m/s, 1~m/s, respectively.
    }
    \label{fig:sim}
\end{figure}
The mechanism described above can be confirmed with numerical simulations of droplet impact. 
The simulations are used to qualitatively reveal the spreading and receding mechanisms on the asymmetric microstructures, which can not be resolved with our experimental setup.
The droplet impact on the asymmetric microstructure is modeled with Navier-Stokes-Cahn-Hilliard equations. The Cahn-Hilliard equation describes the time evolution of the two-phase system based on the diffusion of the chemical potential of the system, whereas the Navier-Stokes equations describe the incompressible flow field. The simulations are carried out in a two-dimensional geometry to keep the computational cost feasible. The droplet radius and the impact velocity are set to $0.125$~mm and $1 - 2$~m/s, respectively. 
Note that the impact velocity, the relative scale of the microstructures to the droplet, and the surface geometry are different from the experiments. 
The details of the simulations are provided in the Supporting Information. 

Figure~\ref{fig:sim}(a) shows snapshots from the simulation of a droplet impacting on the inclined microstructures with $V_0$ = 2~m/s. The droplet is displacement in the direction with the inclination, although the droplet does not detach from the surface. 
Here, the sidewalls of the inclined structures are wetted during the spreading (see the inset in Fig.~\ref{fig:sim}a). The spreading is nearly symmetric until 0.2 ms, then the contact line starts to recede (see Fig.~\ref{fig:sim}b). 
During the retraction in the direction against the inclination, the liquid is arrested on the sidewall of the structures. While the contact line is pinned at the obtuse corner, the liquid phase can not detach from the sidewall (see the inset in Fig.~\ref{fig:sim}a). The apparent contact angle has to decrease below 60 degrees before the contact line detaches from the sidewall (see Fig.~\ref{fig:sim}c). 
On the other hand, during the retraction in the direction with the inclination, the apparent contact angle does not become lower than 80 degrees. Consequently, the retraction is faster in the direction with the inclination than in the direction against the inclination. This difference is responsible for the directional motion in the direction with the inclination.

Moreover, the numerical simulations with different impact velocities are consistent with our experimental observation. Figure~\ref{fig:sim}(d) shows the horizontal displacement of the center of the mass of the droplet with different impact velocities. The larger the impact speed, the faster the horizontal motion becomes.
As seen in Fig.~\ref{fig:sim}(b), the retraction distance is larger for a higher impact velocity.
Here, the longer retraction distance of the contact line is responsible for the stronger effect of the pinning, which leads to the larger displacement.

\subsection{Receding contact angle measurements}

\begin{figure}
    \centering
     \includegraphics[width=0.6\textwidth]{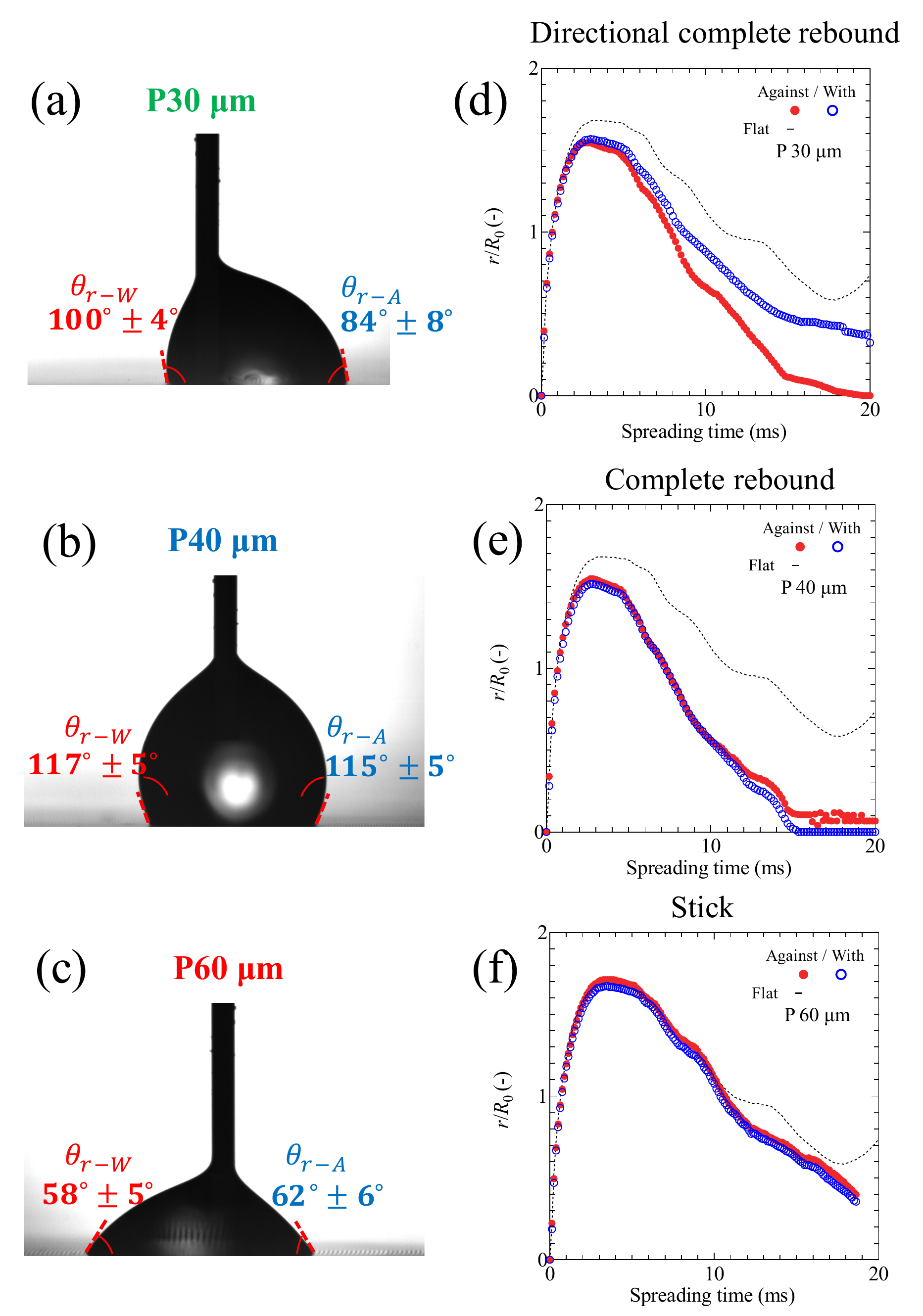}
    \caption{(a-c) Measured receding contact angles on the asymmetric microstructures. The pictures show the typical droplet shapes after the contact line starts to recede. (d-f) Corresponding droplet impact behavior. Contact line position from the initial center of the droplet for the impact speed $V_0$ = 0.64~m/s ($H_0$ = 25~mm). The droplet rebounds directionally in the direction with the inclination on $P= 30~\mathrm{\mu m}$, rebounds vertically on $P= 40~\mathrm{\mu m}$, and sticks on $P= 60~\mathrm{\mu m}$.
    }
    \label{fig:REC_angle_radius}
\end{figure}

Here, we demonstrate that the receding contact angle measured with the sessile drop method is consistent with the droplet impact behavior. 
The receding contact angle in the direction against the inclination ($\theta_{r-A}$ = 84~degrees) is smaller than in the direction with the inclination ($\theta_{r-W}$= 100~degrees) for $P= 30~\mathrm{\mu m}$ (Fig.~\ref{fig:REC_angle_radius}a). 
This is consistent with the fact that the receding speed in the direction against the inclination during the droplet impact is slower (Fig.~\ref{fig:REC_angle_radius}d). This difference leads to directional rebound. 
Meanwhile, the receding angles on the surfaces with $P= 40~\mathrm{\mu m}$ and $60~\mathrm{\mu m}$ 
are similar for the two directions (Fig.~\ref{fig:REC_angle_radius}b, c). The symmetry in the receding contact angle is consistent with the symmetric receding speed during the droplet impact (Fig.~\ref{fig:REC_angle_radius}e, f). 

The pitch in the direction perpendicular to the direction of the inclination of the surface structures is potentially responsible for the difference between $P= 30~\mathrm{\mu m}$ and $P= 40~\mathrm{\mu m}$ where the droplet rebounds in both cases. 
The effect of the pinning described in Fig.~\ref{fig:schematics} is effective only when the pillars are sufficiently dense along the contact line so as for the pinning to be effective enough to delay the receding. 
This implies that the pinning site is dense enough only for $P= 30~\mathrm{\mu m}$ but not for $P= 40~\mathrm{\mu m}$ in our experiments.
Moreover, the mechanisms in Fig.~\ref{fig:schematics} are undermined for "Partial rebound" and "Stick" cases since the grooves between the posts are filled with water. Therefore, the directional behavior is not expected for "Partial rebound" and "Stick" droplets. 

\subsection{Discussion}

 The mechanisms underlying asymmetric droplet rebound elucidate the influence of the surface geometries on rebound behavior.
To realize a directional rebound on arrays of inclined micropillars, three conditions must be fulfilled. First, the receding contact line speeds in the directions with and against the inclination need to be different.
Second, the grooves between the pillars should not be completely wetted. This condition corresponds to the complete rebound regime in the Cassie-Wenzel transition model. Third, the impact velocity should be large enough to deform the droplet and lead to substantial retraction.
To satisfy these three conditions, a surface should have both sufficient pinning corners in the direction perpendicular to the inclination of the surface structures and be hydrophobic enough to induce a rebound.

\begin{figure}[b!]
    \centering
    \includegraphics[width=0.4\textwidth]{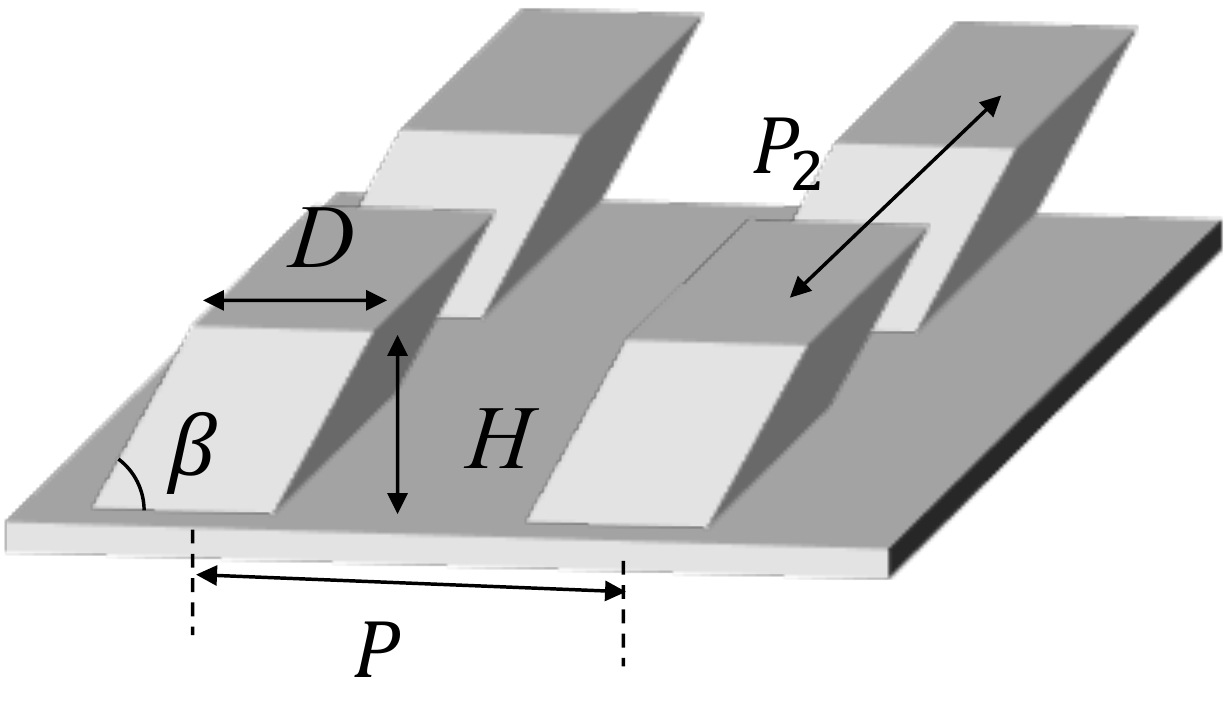}
    \caption{ Another pitch in the direction perpendicular to the inclination of the surface structures.$P_2$
    }
    \label{fig:AnotherP}
\end{figure}

 There is however a trade-off between the density of the pinning corners and the hydrophobicity of the surface. As the pitch decreases ($P \rightarrow 0$), the number of  pinning sites along the contact line increases, but the surface becomes less hydrophobic, as the solid-air ratio increases.
An additional degree of freedom of the surface to enhance the directional rebound, is the pitch $P_2$ in the direction perpendicular to the direction of the inclination (see Fig.~\ref{fig:AnotherP}). Since, it is desirable to increase the number of pinning site while keeping the surface hydrophobic,  structures with $P> P_2$ and a reasonably large static contact angle could enhance the directional rebound.

\begin{figure}[b!]
    \centering
    \includegraphics[width=0.6\textwidth]{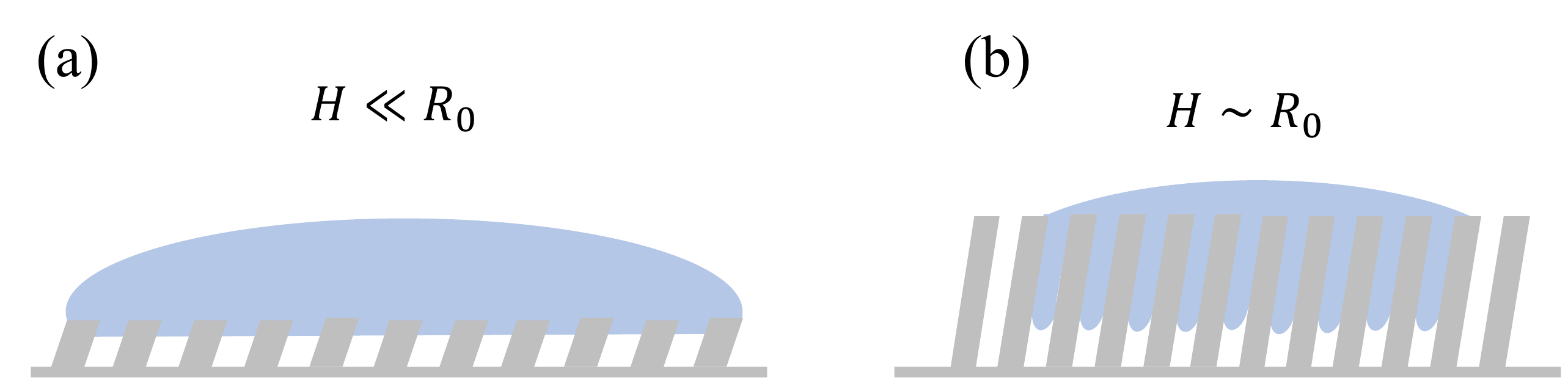}
    \caption{ Schematic descriptiton of the wetting and rebound scenario on microstructures (a) with small height compared to the radius of the droplet (b) with comparable height to the radius of the droplet.
    }
    \label{fig:Height_dif}
\end{figure}

The directional rebound mechanisms proposed in previous studies are  different in certain aspects than in this study. Lee et al. \cite{Lee_ACSn2018} proposed that the stored surface energy between the inclined structures is responsible for the directional rebound. Note that the height of their structures is in the order of 1~mm, whic is two orders of magnitude higher than in this study (and the aspect ratio of the surface dimension is large).
For such surfaces, the droplet can fully penetrate the pillar arrays during impact. A similar large penetration is observed by Li et al. \cite{Li_RESA_2017} The two different situations are schematically shown in Fig.~\ref{fig:Height_dif}.
Lee et al. \cite{Lee_ACSn2018} and Liu et.al.\cite{Liu_ARL_2015} express the change in surface energy during the penetration as $E_s\sim 4\sigma n_p h^2 |\cos\theta_e|$, where $n_p$ is the number of wetted pillars and $h$ is the penetration depth. The surface energy is expected to transform to the kinetic energy of the bouncing droplet, assuming viscous dissipation is negligible. The eject velocity given by the surface energy is directed into the direction of the inclination, led by capillary forces.
Assuming $n_p \sim (R_0/P)^2$ and $h \sim H$, $E_s \sim 4\sigma R_0^2 (H/P)^2 |\cos\theta_e|$. 
Since $E_s$ is proportional to the square of the ratio of the height and the pitch $H/P$, the change of surface energy by the penetration could be small for our surface.
For example, an order of magnitude estimation for a droplet with a initial radius of 1~mm, $E_s \sim O(10^{-7})$ for our structures with $H/P \sim 1$ while $E_s \sim O(10^{-5})$ for Lee et al. with $H/P \sim 10$ \cite{Lee_ACSn2018}. 
This is equal to the kinetic energy of the droplet with velocity $\sim$ 0.1 m/s for our structures and $\sim$ 1 m/s for Lee \textit{et al.}
Therefore, the change of surface energy from the penetration is insignificant for surface roughness with an aspect ratio of $\sim$ 1 and $H \ll R_0$. Instead, the difference in the receding contact line speed is responsible for the directional rebound instead.

The different mechanisms may explain the differences in the horizontal velocity and the displacement. Here, we compare the horizontal velocity and displacement distance on the inclined microstructures with $\beta \sim$ 60 degrees in the literature. 
For a smaller structure where the rebound mechanisms are described in Fig.~\ref{fig:Height_dif}(a), smaller displacement and velocity are observed. 
Li \textit{et al.} \cite{PLi_ACSAMI_2021} reported that on inclined cone structures with $H\sim 300~\mathrm{\mu m}$ and $P\sim 300-400~\mathrm{\mu m}$ , the displacement is 0.7~mm for $V_0$= 0.71~m/s and 2.0~mm for $V_0$= 1.73~m/s. The horizontal rebound distance is similar to this study. 

On the other hand, for a very tall structure with $H > 1~\mathrm{mm}$ and $P < 500~\mathrm{\mu m}$ , where the rebound may follow the scenario in Fig.~\ref{fig:Height_dif}(b), a larger rebound velocity is observed.
 Lee \textit{et al.} \cite{Lee_ACSn2018} reported the horizontal velocity and displacement on thin-spike structures. The horizontal velocity is 0.09~m/s and displacement distance is 9.1~mm for $V_0$= 1.13~m/s.
Similarly, Li \textit{et al.} \cite{Li_RESA_2017} reported a large horizontal velocity and horizontal displacement of 0.062~m/s and 5.5~mm, but the impact velocity information is missing. 
The difference in the mechanisms could be responsible for the larger displacement since the surfaces described in Fig.~\ref{fig:Height_dif}(b) are capable of harnessing droplets with larger impact speed and therefore able to store larger surface energy before the rebound. It is worth noting that Lee \textit{et .al.} \cite{Lee_ACSn2018} also pointed out that the bouncing mechanisms on superhydrophobic nanostructures and hairy spike structures are different. While the droplet rebounds after the full retraction on the superhydrophobic nanostructures, the droplet rebounds by upward capillary forces on their hairy spike arrays. 

\section{Conclusions}
We studied the droplet impact on asymmetric microstructures. 
Directional rebound was observed only for dense microstructures and at high impact speeds. 
The retraction phase and the detailed wetting of the sidewalls of the inclined structures govern the rebound.
The wetting of the sidewall leads to a slower receding speed in the direction against the inclination.
The contact line can be pinned at the obtuse corner when receding in the direction against the inclination, while in the other direction the contact line recedes continously. 
The receding contact angles on the asymmetric pillar structures correlate with the droplet rebound behavior. The directional rebound is found only on surfaces with asymmetric receding contact angles in the direction of the pillar inclination.
Numerical simulations provide further detailed visualizations of the two phase interface near the pillars and confirm our experimental observations.
We hope that insights gained in this study will be useful for tuning surface structures for directional transport of liquid drops.



\section{Acknowledgements}
This work was supported by the Swedish Research Council (VR 2015-04019) 
and by the Swedish Foundation of Strategic research (SSF-FFL6). 
The authors thank Rohan Kulkarni at Royal Institute of Technology for his kind help with SEM measurements.
\begin{suppinfo}
The detailed numerical methods.
Video animations of a simulated droplet impact.
\end{suppinfo}

\bibliography{achemso-demo}

\end{document}